\documentclass[preprint,showpacs,preprintnumbers,amsmath,amssymb]{revtex4}

\usepackage{graphicx}
\usepackage{dcolumn}
\usepackage{bm}

\begin{document}


\title{Is the universe rotating?}

\author{Shi Chun, Su}
 \email{scsu@phy.cuhk.edu.hk}
\author{M.-C., Chu}%
 \email{mcchu@phy.cuhk.edu.hk}
\affiliation{%
Department of Physics and Institute of Theoretical Physics, The Chinese University of Hong Kong, Shatin, Hong Kong
}%

\date{\today}

\begin{abstract}
Models of a rotating universe have been studied widely since G{\"o}del \cite{1}, who showed an example that is consistent with General Relativity (GR). By now, the possibility of a rotating universe has been discussed comprehensively in the framework of some types of Bianchi's models, such as Type V, VII and IX \cite{2,3}, and different approaches have been proposed to constrain the rotation. Recent discoveries of some non-Gaussian properties of the Cosmic Microwave Background Anisotropies (CMBA) \cite{nG1,nG2,nG3,nG4,nG5,nG6,nG7}, such as the suppression of the quadrupole and the alignment of some multipoles draw attention to some Bianchi models with rotation \cite{bi1,bi2}. However, cosmological data, such as those of the CMBA, strongly prefer a homogeneous and isotropic model. Therefore, it is of interest to discuss the rotation of the universe as a perturbation of the Robertson-Walker metric, to constrain the rotating speed by cosmological data and to discuss whether it could be the origin of the non-Gaussian properties of the CMBA mentioned  above. Here, we derive the general form of the metric (up to 2nd-order perturbations) which is compatible with the rotation perturbation in a flat $\Lambda$-CDM universe. By comparing the 2nd-order Sachs-Wolfe effect \cite{4,5,6,7,8} due to rotation with the CMBA data, we constrain the angular speed of the rotation to be less than $10^{-9}$ rad yr$^{-1}$ at the last scattering surface. This provides the first constraint on the shear-free rotation of a $\Lambda$CDM universe.

\end{abstract}

\pacs{Unknown}
\maketitle

\section{\label{sec:level1}Introduction}

In contrast to most astronomical objects (such as planets, stars and galaxies), the universe as a whole is usually considered to be non-rotational. However, the possibility that the universe rotates should not be ignored, since solutions of GR corresponding to a rotating universe have been found \cite{1,godelBarrow,2,3} indicating that a global rotation is physically allowed. Although it is widely assumed that galaxies align randomly and have zero net angular momentum, there have been many investigations on the general alignment of galaxies. Some even claim that a weak alignment of galaxies does exist\cite{alignm1,alignm2}. Such an alignment may be used to explain\cite{Longo} the recently discovered non-Gaussian properties\cite{nG1,nG2,nG3,nG4,nG5,nG6,nG7} of the CMBA. Furthermore, Jaffe et al. \cite{bi1,bi2} suggest that the Bianchi Type VII$_\text{h}$ model with a global rotation may be used to explain some anomalies of the CMBA. In addition, the existence of a global rotation may contradict the inflationary model of the early universe \cite{inflation1,inflation2,inflation3} and Mach's Principle \cite{Mach1,Mach2}. Rotational perturbations may also be used to determine whether or not the universe is open or closed \cite{3,Barrow3}. Therefore, the study of global rotation is of interest in many different aspects of cosmology, and constraint of the rotation speed of the universe is important. 

The most popular approach to constrain the magnitude of the global rotation speed is to make use of data from the CMBA because of their precision. Most discussions \cite{2,3,COBB,WMAAA} have focused on homogeneous cosmological models, i.e. Bianchi models \cite{bianchi}. To be consistent with obeservations, only Bianchi Type V, VII$_0$, VII$_\text{h}$ and IX models, which include the Robertson-Walker model as a special case, are considered. The constraints of the global rotation speed obtained depend on the parameters of the models. Besides, shear and vorticity are inseparable in these works \cite{Shear}, i.e. zero shear automatically implies zero vorticity.

There are many other approaches to constrain the global rotation. Based on the idea that a global rotation induces a total net spin of galaxies, the global rotation can be limited \cite{JM1}. Moreover, empirical relations between angular momenta and mass of galaxies/clusters, such as $J\sim M^{5/3}$ for spiral galaxies and $J\sim M^{2}$ for clusters can be explained by the global rotation \cite{JM1,JM2}. The acceleration caused by the global rotation may be used to explain parts of the accelerating expansion of our universe, and thus the global rotation can be constrained by Supernova Type Ia data \cite{SNI}. Recently, some studies of the CMB polarization induced by the global rotation are published \cite{Polar} providing potential constraints in the future.

To develop a model that preserves the homogeneity and isotropy of the mean CMB, we study the rotation of the universe as a perturbation in the Robertson-Walker framework with a cosmological constant in this paper. Unlike the Bianchi models, such an approach allows to have non-zero rotation but trivial shear. Since the global rotation does not have any influences on the 1st-order Sachs-Wolfe effect (SW effect), we need to calculate the metric up to 2nd-order perturbations and the 2nd-order SW effect. Then, we will constrain the angular speed of the rotation using recent data on CMBA \cite{nasa}. Our model is inhomogeneous with an axial symmetry in general. The global rotation in our model is not only time-dependent but also radial-dependent.

\section{Solutions of the Einstein field equations with rotational perturbations}
The line element of a flat rotational universe possesses an axial symmetry and can be written in the form of \cite{metri}
\begin{eqnarray}
ds^2&=&g_{\mu\nu}dx^{\mu}dx^{\nu}\\
&=& a^2(\eta)\{[1-f(r,\eta)]d\eta^2-[1-h(r,\eta)]dr^2-[1-h(r,\eta)]r^2d\theta^2-
[1-k(r,\eta)]dz^2\nonumber\\
&~&+2r^2a(\eta)\Omega(r,\eta)d\theta d\eta\},
\end{eqnarray}
where $\mu$ and $\nu=0,1,2,3$, $x^{\mu}=\{\eta,r,\theta,z\}$, $\eta$ is the conformal time defined by $dt=a(\eta)d\eta$ with $t$ the cosmological time, $r$, $\theta$ and $z$ are the cylindrical coordinates in the comoving frame of the universe, $\hat{z}$ is the axis of rotation, $a(\eta)$ is the scale factor of the universe with $a(\eta=0)=1$ at the present time, $\Omega(r,\eta)$ is the angular velocity of the metric observed from an inertial frame whose origin is on the rotational axis, and $f(r,\eta)$, $h(r,\eta)$ and $k(r,\eta)$ are the perturbations on the (0, 0), (1, 1), (2, 2), (3, 3)-components of the metric due to the rotation. Because of the cylindrical symmetry, the perturbation functions due to the rotation are also independent of $\theta$ and $z$.

Here, we assume that the norm of $g_{\mu\nu}$$-$$_0g_{\mu\nu}$, where $_0g_{\mu\nu}$ is the unperturbed metric, is much smaller than that of $g_{\mu\nu}$. Explicitly, we assume that the rotation is slow, so that $ra(\eta)\Omega(r,\eta)\ll 1$, and we can think of $\Omega_{rot}\equiv\max\Omega(r,\eta)$ for $r$, $\eta$ within the last scattering surface as the perturbation parameter. By parity consideration, we can see that $\Omega(r,\eta)$ is composed of only odd powers of $\Omega_{rot}$, whereas $f(r,\eta)$, $h(r,\eta)$ and $k(r,\eta)$, being density and pressure perturbations, only even powers. Since we are interested only up to second-order perturbations, we will consider $\Omega(r,\eta)$ to be first-order and $f(r,\eta)$, $h(r,\eta)$ and $k(r,\eta)$ to be second-order. The metric Eq.~(1.1) in Ref.~\cite{1storder} will be recovered if we truncate ours up to the first-order. Since the effect of the rotation on the CMBA is independent of the parity, we expect that the SW effect due to rotation occurs in even orders of $\Omega_{rot}$ only.

The Einstein Field equations (EFEs) for a universe with cosmological constant $\Lambda$ are 
\begin{eqnarray}
R_{\mu\nu}-\frac{1}{2}g_{\mu\nu}R+\Lambda g_{\mu\nu}=G_{\mu\nu}+\Lambda g_{\mu\nu}&=&8\pi T_{\mu\nu},
\end{eqnarray}
where $T_{\mu\nu}=(\rho+P)u_{\mu}u_{\nu}-Pg_{\mu\nu}$ is the stress-energy tensor for a perfect fluid, $R_{\mu\nu}$ is the Ricci curvature tensor, $R$ is the scalar curvature, $\rho$ is the mass-energy density, $P$ is the pressure and $u^{\mu}=a^{-1}(\eta)dx^{\mu}/d\eta$ is the four-velocity of the fluid in the comoving frame. Here, we set $G=c=1$.

If $\Omega(r,\eta)=\Omega(\eta)$, the universe is homogeneous and $\Omega(\eta)$ expresses the angular velocity of the universe observed anywhere in the comoving frame. Otherwise, the universe is inhomogeneous and the observer at the rotating axis passing through the origin is distinct. To solve Eq.~(3) up to 2nd-order in $\Omega_{rot}$, we expand all quantities:
\begin{eqnarray}
g_{\mu\nu}&=&_0g_{\mu\nu}+{_1g_{\mu\nu}}+{_2g_{\mu\nu}},\\
\rho&=&_0\rho+{_1\rho}+{_2\rho},\\
P&=&_0 P+{_1 P}+{_2 P},\\
u^{\mu}&=&_0u^{\mu}+{_1u^{\mu}}+{_2u^{\mu}},
\end{eqnarray}
where the subscripts indicate the corresponding orders of perturbations.

The zeroth-order EFEs give rise to the standard Friedmann equations:
\begin{eqnarray}
\frac{3\dot{a}^2(\eta)}{a^2(\eta)}+\Lambda a^2(\eta)&=&8\pi a^2(\eta) _0\rho(\eta),\\
-\Lambda a^2(\eta)+\frac{\dot{a}^2(\eta)}{a^2(\eta)}-\frac{2\ddot{a}(\eta)}{a(\eta)}&=&8\pi a^2(\eta)_0P(\eta).
\end{eqnarray}
Once the equation of state (EOS) of the fluid is given, we can determine the scale factor $a(\eta)$, the density $_0\rho(\eta)$ and the pressure $_0P(\eta)$ with the equations above. In this paper, we consider a universe with $_0P(\eta)=0$. However, the following formalism can be applied to any fluid with a specified EOS.

From the temporal-spatial EFEs and the condition $u_{\mu}u^{\mu}=1$, we have
$_0u_{\mu}(\eta)=(a(\eta),0,0,0)$ and $_1u_0(r,\eta)=0$.
The first-order EFEs then give:
\begin{eqnarray}
_1\rho(r,\eta)&=&0,~_1P(r,\eta)=0,~_1u_1(r,\eta)=0,~_1u_3(r,\eta)=0,\\
_1u_2(r,\eta)&=&r^2a^2(\eta)[\Omega(r,\eta)-{_1u^2(r,\eta)}]=\frac{a^4(\eta)}{8\dot{a}^2(\eta)- 4a(\eta)\ddot{a}(\eta)}[3r\Omega '(r,\eta)+r^2\Omega ''(r,\eta)],\\
0&=&3\dot{a}(\eta)\Omega '(r,\eta)+a(\eta)\dot{\Omega}'(r,\eta),
\end{eqnarray}
where the dots refer to derivatives with respect to the conformal time $\eta$, and primes mean derivatives with respect to $r$.

As seen from the equations above, a 1st-order rotational perturbation cannot generate 1st-order perturbations of the mass-energy density and pressure. This is expected because $\rho$ and $P$ should be unchanged under the inversion of the rotation. For the same reason, $_1u_1(r,\eta)={_1u_3(r,\eta)}=0$. From Eq.~(11), we see that if $\Omega(r,\eta)$ is independent of $r$, then $_1u^2(r,\eta)=\Omega(\eta)$. That is, the fluid in the universe rotates with the metric at the same pace. Nevertheless, an $r$-dependent $\Omega(r,\eta)$ allows us to discuss the centrifugal force for the universe as that discussed in Ref.~\cite{star1,star2} for relativistic stars. Eq.~(12) implies that $\Omega(r,\eta)$ must be in the form of $a^{-3}(\eta)A(r)+B(\eta)$, where $A(r)$ and $B(\eta)$ are arbitrary functions. Moreover, if the fluid is viscous, the R.H.S. of Eq.~(12) will be equal to the 1st-order shear term of $8\pi T_{r\theta}$ and this will free $\Omega(r,\eta)$ from the form above.

Without loss of generality, we perform the following transformations: 
\begin{eqnarray}
f(r,\eta)&=&r^2a^2(\eta)\Omega^2(r,\eta)-k(r,\eta)-T(r,\eta),\\
h(r,\eta)&=&k(r,\eta)-L(r,\eta),
\end{eqnarray}
where $T(r,\eta)$ and $L(r,\eta)$ are arbitrary functions depending on $f(r,\eta)$ and $h(r,\eta)$. The first term of $f(r,\eta)$ comes from the transformation  $d\theta\rightarrow d\theta-a(\eta)\Omega(r,\eta)d\eta$. Using these transformations to formulate the second-order EFEs, we find that
\begin{eqnarray}
_2u_2(r,\eta)&=&_2u_3(r,\eta)=0,\\
_2u_0(r,\eta)&=&\frac{a(\eta)}{2}\left\{-2_1u_2\Omega(r,\eta)+\frac{[_1u_2(r,\eta)]^2}{r^2a^2(\eta)}
+k(r,\eta)+T(r,\eta)\right\},\\
_2u_1(r,\eta)&=&\frac{-a^2(\eta)}{8\dot{a}^2(\eta)-4a(\eta)\ddot{a}(\eta)}
\{-\dot{a}(\eta)[2k'(r,\eta)+2T'(r,\eta)]\nonumber\\
&~&+a(\eta)[-2\dot{k}'(r,\eta)+\dot{L}'(r,\eta)]\},\\
_2\rho(r,\eta)&=&-\frac{1}{32\pi a^2(\eta)}\{-4\Lambda a^2(\eta)[k(r,\eta)+T(r,\eta)]
+\frac{4\dot{a}(\eta)[3\dot{k}(r,\eta)-2\dot{L}(r,\eta)]}{a(\eta)}\nonumber\\
&~&-\frac{2[2k'(r,\eta)-L'(r,\eta)]}{r}+12ra^2(\eta)\Omega(r,\eta)\Omega '(r,\eta)+r^2a^2(\eta)\Omega '^2(r,\eta)-4k''(r,\eta)\nonumber\\
&~&+2L''(r,\eta)+4r^2a^2(\eta)\Omega(r,\eta)\Omega ''(r,\eta)
+4\left[\Lambda a^2(\eta)+\frac{3\dot{a}^2(\eta)}{a^2(\eta)}\right]\{k(r,\eta)
+T(r,\eta)\nonumber\\
&~&-r^2a^2(\eta)\Omega^2(r,\eta)+\left[ra(\eta)\Omega(r,\eta)-
\frac{_1u_2(r,\eta)}{ra(\eta)}\right]^2\}\},\\
_2P(r,\eta)&=&\frac{1}{32\pi ra^4(\eta)}\{
4ra(\eta)[3\dot{a}(\eta)\dot{k}(r,\eta)
-\dot{a}(\eta)\dot{L}(r,\eta)+\dot{a}(\eta)\dot{T}(r,\eta)]+
2a^2(\eta)[2r\ddot{k}(r,\eta)\nonumber\\
&~&-r\ddot{L}(r,\eta)+T'(r,\eta)]
+ra^4(\eta)[-4\Lambda k(r,\eta)-4\Lambda T(r,\eta)+r^2\Omega '^2(r,\eta)]\},\\
_2P(r,\eta)&=&\frac{1}{32\pi r^2a^6(\eta)}\{-12\dot{a}^2(\eta)[_1u_2(r,\eta)]^2+4r^2a^3(\eta)[3\dot{a}(\eta)\dot{k}(r,\eta)
-\dot{a}(\eta)\dot{L}(r,\eta)\nonumber\\
&~&+\dot{a}(\eta)\dot{T}(r,\eta)]-r^2a^6(\eta)[4\Lambda k(r,\eta)+4\Lambda T(r,\eta)+3r^2\Omega '^2(r,\eta)]\nonumber\\
&~&+a^4(\eta)\{-4\Lambda[_1u_2(r,\eta)]^2+2r^2[2\ddot{k}(r,\eta)-\ddot{L}(r,\eta)+T''(r,\eta)]\}\},\\
_2P(r,\eta)&=&\frac{1}{32\pi ra^4(\eta)}\{
4ra(\eta)\dot{a}(\eta)[3\dot{k}(r,\eta)-2\dot{L}(r,\eta)+\dot{T}(r,\eta)]
-ra^4(\eta)[4\Lambda k(r,\eta)\nonumber\\
&~&+4\Lambda T(r,\eta)+r^2\Omega '^2(r,\eta)]
+2a^2(\eta)[2r\ddot{k}(r,\eta)-2r\ddot{L}(r,\eta)+L'(r,\eta)+T'(r,\eta)\nonumber\\
&~&+rL''(r,\eta)+rT''(r,\eta)]\}.
\end{eqnarray}

Eqs.~(19)-(21) are three different expressions for $_2P(r,\eta)$ derived by the three 2nd-order spatial-spatial EFEs. Eq.~(15) is expected by considering the symmetries of $\theta$(odd)-, $z$(even)-components of the four-velocity $u^{\mu}$ under the inversion of the rotation. $_2u^1(r,\eta)=- _2u_1(r,\eta)/a^2(\eta)$, which is non-zero in general and corresponds to the dynamical changes of $_2\rho(r,\eta)$ and $_2P(r,\eta)$ for an $r$-dependent rotational speed. In order to calculate these 2nd-order perturbations, we have to find the solutions of $k(r,\eta)$, $L(r,\eta)$ and $T(r,\eta)$. Since the pressure is the same along different directions at one point, Eqs.~(19)-(21) are equivalent. Substracting Eqs.~(19)-(21) from each other leads to two equations for solving $L(r,\eta)$ and $T(r,\eta)$ when $\Omega(r,\eta)$ is specified while $k(r,\eta)$ is regarded as an arbitrary function independent of the rotation. The detailed derivations of these solutions are shown in APPENDIX I.

\section{The Sachs-Wolfe effects up to second order}
As the SW effect is invariant under the inversion of the rotation, the first non-zero SW effect due to rotation occurs in 2nd-order perturbations. The general formalism of the 2nd-oder SW effect has been comprehensively discussed. In the following, we will make use of the ideas in \cite{5,6,7,8} and derive the 2nd-order SW effect of a rotating universe.

The Cosmic Microwave Background (CMB) temperature observed at the origin towards a direction $\hat{e}$ can be written as 
\begin{eqnarray}
T_O(\hat{e})=\frac{\omega_O}{\omega_{\epsilon}}T_{\epsilon}(\vec{p},\hat{d}),
\end{eqnarray}
where $\omega=-a^{-2}g_{\mu\nu}u^{\mu}k^{\nu}$, the subscripts ($O$ and $\epsilon$) denoting the origin and the last scattering hypersurface (LSH) respectively, $u^{\mu}=a^{-1}(\eta)dx^{\mu}/d\eta$ is the four-velocity of the fluid in the comoving frame, $k^{\nu}=dx^{\nu}/d\lambda$ is the wave vector of a light ray in the conformal metric with an affine parameter $\lambda$, $T_{\epsilon}(\vec{p},\hat{d})$ is the temperature measured at the point $\vec{p}$ on the LSH, and $\hat{d}$ is the direction of the light (passing through the point $\vec{p}$) observed at the origin.

We show in Appendix II that the 1st-order SW effect due to rotation is zero and the 2nd-order SW effect is 

\begin{eqnarray}
&~&\frac{\delta T_{\epsilon}}{T_{\epsilon}}
=\left[-\frac{_2u_1(r_{\lambda},\eta_{\lambda})}{a(\eta_{\lambda})}\sin\phi
+\frac{_2u_0(r_{\lambda},\eta_{\lambda})}{a(\eta_{\lambda})}+_2k^0(\lambda)
+\frac{_1u_2(r_{\lambda},\eta_{\lambda})}{a(\eta_{\lambda})}{_1k^2(\lambda)}
\right]|_{\eta_{\epsilon}}^{\eta_0}\nonumber\\
&=&\frac{\Omega_{\Lambda}\sin\phi}{2\Lambda(1-\Omega_{\Lambda})}
[2\dot{a}(\eta_{\epsilon})T'(r_{\epsilon},\eta_{\epsilon})-
a(\eta_{\epsilon})\dot{L}'(r_{\epsilon},\eta_{\epsilon})]\nonumber\\
&~&-\frac{1}{2}\left\{\frac{\Omega_{\Lambda}^2a^4(\eta_{\epsilon})}{4\Lambda^2(1-\Omega_{\Lambda})^2}
[3\Omega '(r_{\epsilon},\eta_{\epsilon})+r_{\epsilon}\Omega ''(r_{\epsilon},\eta_{\epsilon})]^2+T(r_{\epsilon},\eta_{\epsilon})\right\}\nonumber\\
&~&+\int_{\eta_{\epsilon}}^0\left[-\frac{\dot{T}(-\lambda\sin\phi,\lambda)}{2}+
T'(-\lambda\sin\phi,\lambda)\sin\phi-\frac{\dot{L}(-\lambda\sin\phi,\lambda)}{2}\sin^2\phi\right]d\lambda,
\end{eqnarray}
where $r_{\epsilon}=-\eta_{\epsilon}\sin\phi$ and $\eta_{\epsilon}$ denotes the conformal time of the last scattering.

Eq.~(23) determines the CMBA produced by the rotation of the universe once $\Omega(r,\eta)$ is specified. As an example, we consider the simplest case -- 
stationary homogeneous rotation (i.e. $\Omega(r,\eta)=B(\eta)$, $B(\eta)$ is an arbitrary function). Then, we have $f(r,\eta)=C\Omega_{rot}^2r^2$ and $h(r,\eta)=0$ where $C$ is a constant. It is straight-forward to find that 
\begin{eqnarray}
\frac{\delta T_{\epsilon}}{T_{\epsilon}}&=&0.
\end{eqnarray}
To explain this, we recall that $_1u^2(r,\eta)=\Omega(r,\eta)$ if $\Omega(r,\eta)$ is independent of $r$, which means that the fluid is rotating with the same phase as the metric. Therefore, the effect of the rotating metric cancels the relativistic Doppler effect caused by the sources rotating in a stationary metric. 

We make use of the previous example, i.e. $\Omega(r,\eta)=\Omega_{rot} a^3(\eta_{\epsilon})r^2/(r_{\epsilon}^2a^3(\eta))$ with $\alpha=a^3(\eta_{\epsilon})/r_{\epsilon}^3$ in Eqs.~(A.10)-(A.18), to constrain the rotation of the universe. Using Eqs.~(23), (A.10), (A.16)-(A.18), we expand the CMBA as
\begin{eqnarray}
\frac{\delta T_{\epsilon}}{T_{\epsilon}}&=&
a_2\sin^2\phi+a_4\sin^4\phi+a_6\sin^6\phi\nonumber\\
&=&A_0Y_0^0(\phi,\theta)+A_2Y_2^0(\phi,\theta)+A_4Y_4^0(\phi,\theta)+A_6Y_6^0(\phi,\theta).
\end{eqnarray}
The values of $A_n$'s are listed in Table 1. \\
\vspace{4ex}
\textbf{Table 1}\\
\begin{tabular}{lllll}
	\hline
$n$ &  $~~~10^{-28} A_n/c^2\Omega_{rot}^2$ $(s^2)$ \\
	\hline\hline
$0$  & ~~~-6.14188 \\
$2$  & ~~~~4.57532 \\
$4$  & ~~~-1.67194 \\
$6$  & ~~~~0.25710 \\
	\hline
\end{tabular}\\

We notice that the spherical harmonic expansion has non-zero coefficients only when $m=0$ and even $l$ for which $Y_l^m(\phi,\theta)$ has cylindrical and parity symmetries. The result is unlikely to be related to the `Axis of Evil' \cite{nG1,nG2,nG3,nG4,nG5,nG6,nG7}, a preferred direction of several low multipoles (especially quadrupole and octopole). However, in the general case, when we are located off the rotational axis, the cylindrical symmetry is broken and non-zero coefficients for other multipoles are allowed. For example, the CMBA on the two sides of the rotation axis will be affected differently by the rotation in general. Such an asymmetric effect enhances the dipole moment of the CMBA. Thus, its potential to explain the `Axis of Evil' cannot be eliminated without further study. 

With $\Omega_{\Lambda}=0.742$, $H_0=71.9$km/s/Mpc and $A_n\sim 10^{-5}$, we constrain $\Omega_{rot}$ to be less than $\sim 6 \times 10^{-26}$ m in SI unit. That is, $\Omega(r_{\epsilon},\eta_{\epsilon})$ is less than $\sim 10^{-9}$ rad yr$^{-1}$ in usual unit at the last scattering surface. Some CMBA maps generated with the rotation of the universe are shown in Fig.~1 as examples. Nevertheless, our result can be regarded as the first constraint of the rotation of a $\Lambda$CDM universe.

In Fig.~2, some normalized 2nd-order perturbed quantities along the light path of the last scattered photons are plotted as a function of $ra(\eta)\Omega(r,\eta)$ with $\Omega(r,\eta)=\Omega_{rot} a^3(\eta_{\epsilon})r^2/(r_{\epsilon}^2a^3(\eta))$ and $\Omega_{rot}\sim 6 \times 10^{-26}$ m. As expected, the perturbed quantities increase with the rotating speed. In Fig.~3, the angular velocity of matter $_1u^2(r_,\eta)$ and its difference from that of the metric along the light path of the last scattered photons are plotted against time. We can see that the angular velocity of matter can be negative while the rotation speed of the universe is always positive. Because of the $r$-dependence of $\Omega(r,\eta)$, the angular velocity of matter can be different from that of the metric in general as indicated in Eq.~(11). These quantities are useful for studying the frame-dragging of the universe in the future. The distributions of the 2nd-order perturbed densities of matter are shown at two different times in Fig.~4. As shown in Fig.~2, $_2u^1(r,\eta)$ is always positive, which means that matter is moving away from the rotating axis and hence the density is expected to be decreasing with time (shown in Fig.~4).

\begin{figure}[hp]
\centering
\includegraphics[width=9cm, height=5cm]{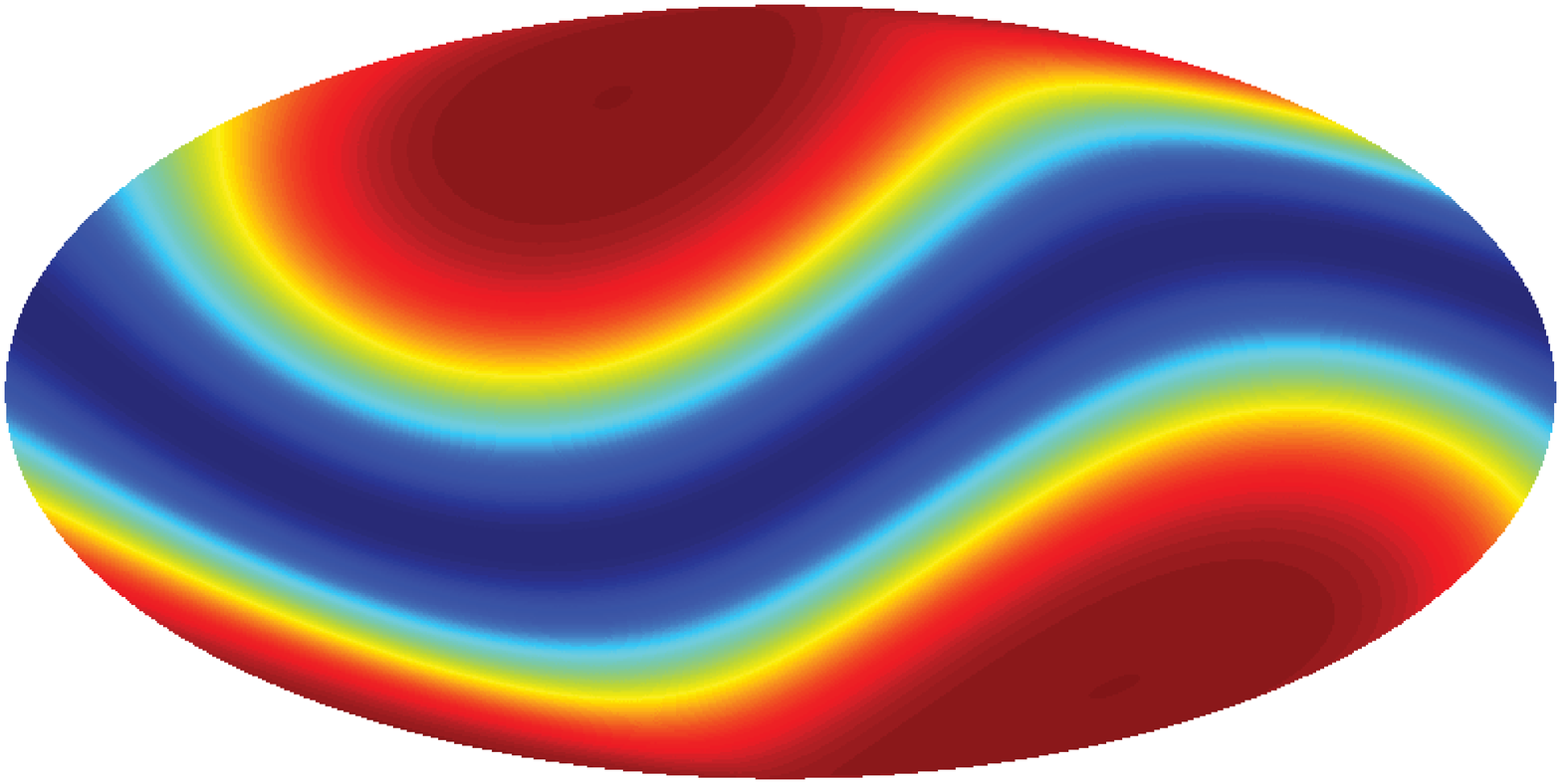}
\includegraphics[width=9cm, height=5cm]{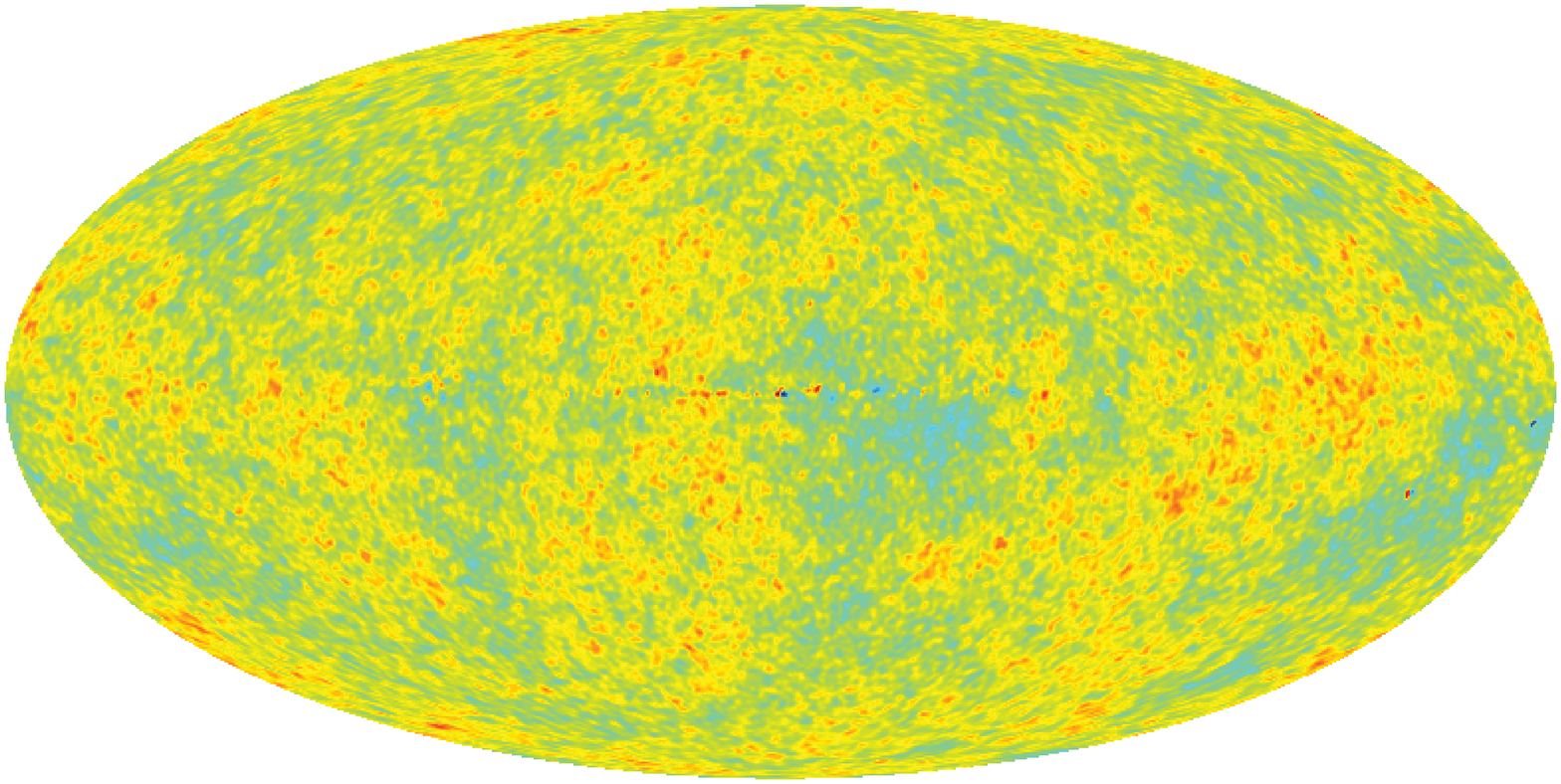}
\includegraphics[width=9cm, height=5cm]{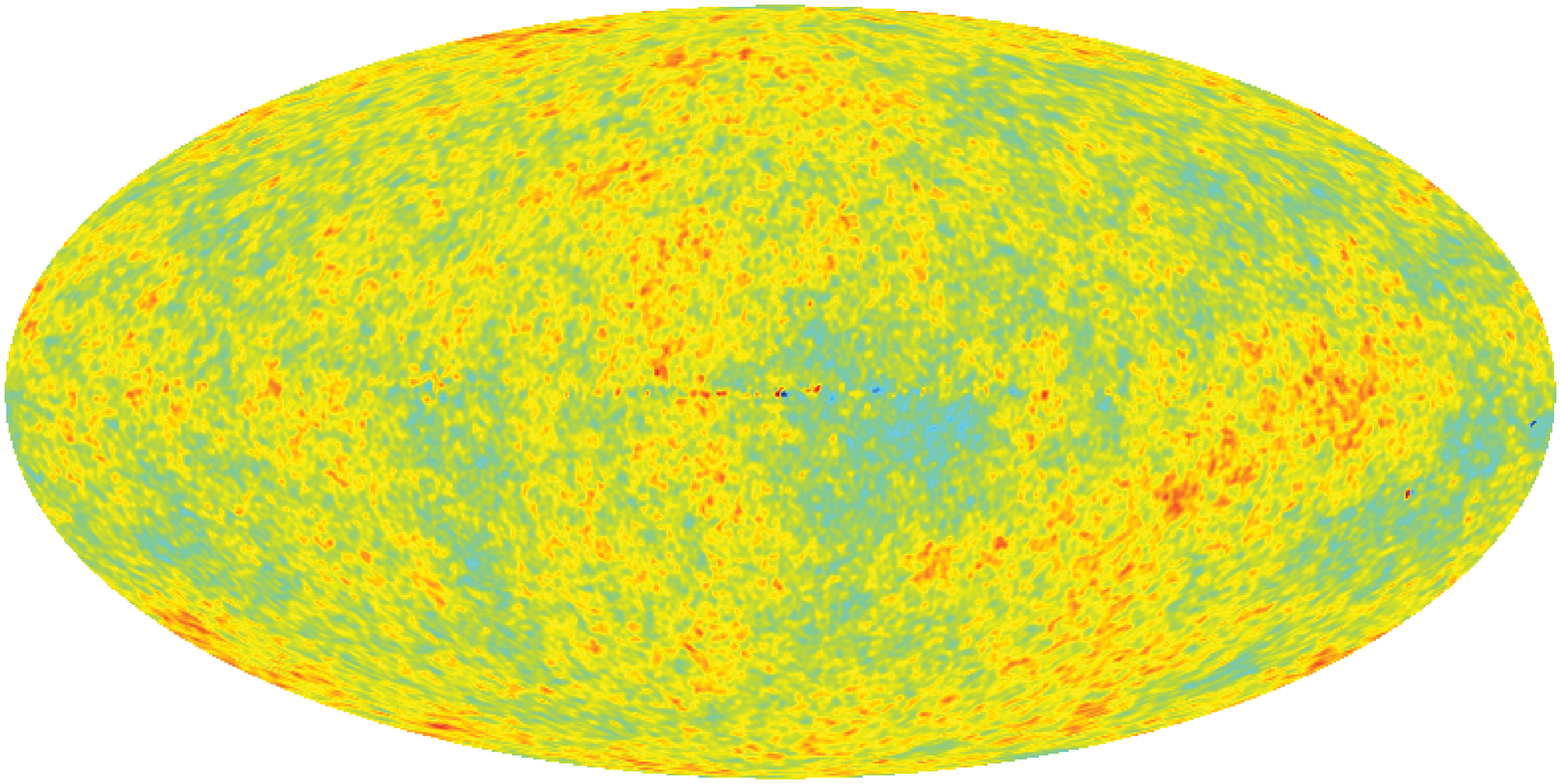}
\caption{The top map shows the effects of the rotation on $\delta T_{\epsilon}/T_{\epsilon}$ under the Mollweide projection with the z-direction pointing to $(b,l)=(60^{\circ},120^{\circ})$ with $\Omega_{rot}\sim 6 \times 10^{-26}$ m, which is the maximum allowed by current CMB data. The middle map shows the original 5-year WMAP map \cite{nasa}, and the bottom map is a combined map of the two above.}
\label{fig:}
\end{figure}

\begin{figure}[hp]
\centering
\includegraphics[width=16cm, height=12cm]{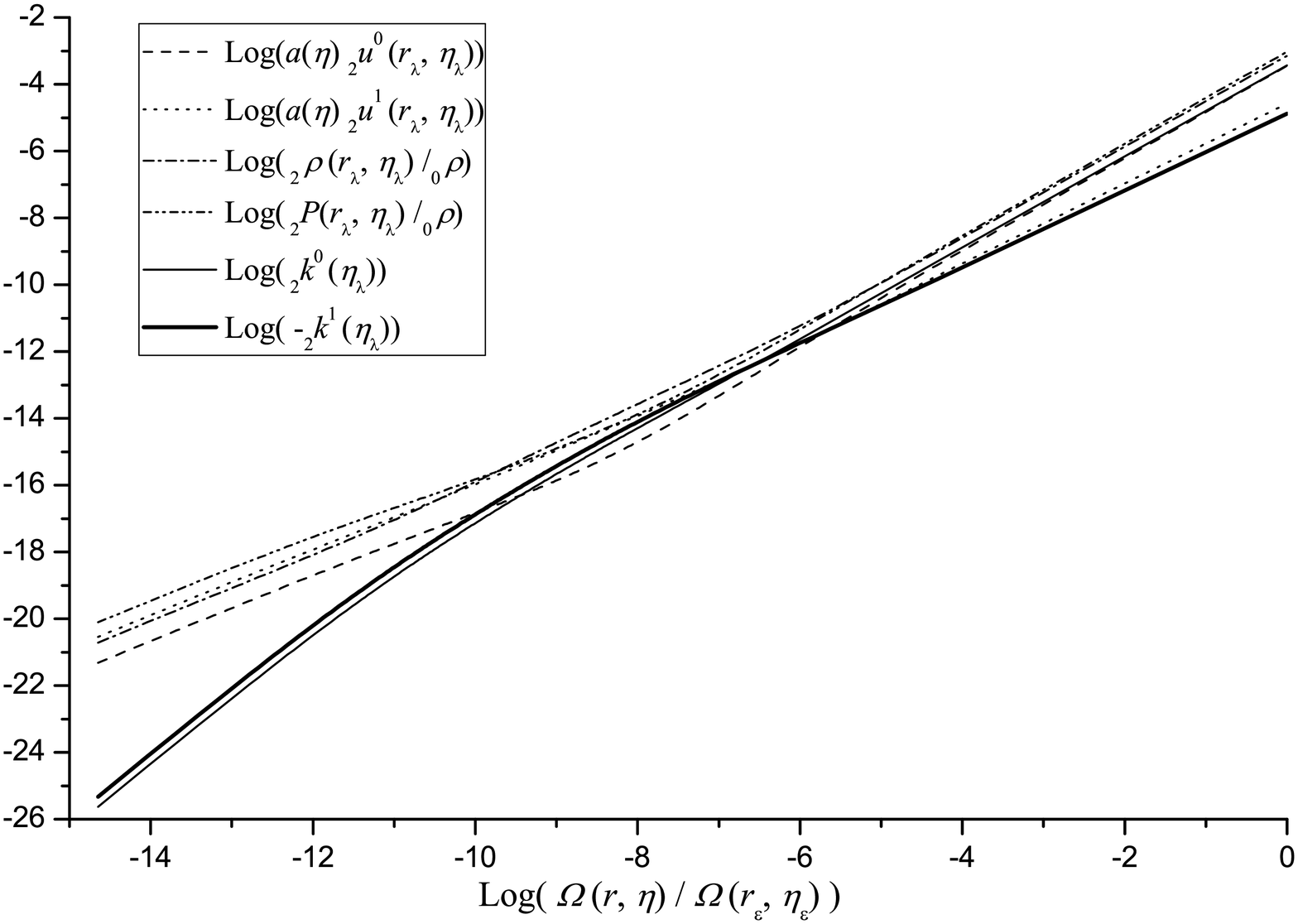}
\caption{Some 2nd-order perturbed quantities are shown against $ra(\eta)\Omega(r,\eta)$ along the light paths of the last-scattered photons, where $r_{\lambda}=-\lambda\sin\phi$ and $\eta_{\lambda}=\lambda$. We set $\phi=\pi/2$.}
\label{fig:}
\end{figure}

\begin{figure}[hp]
\centering
\includegraphics[width=16cm, height=12cm]{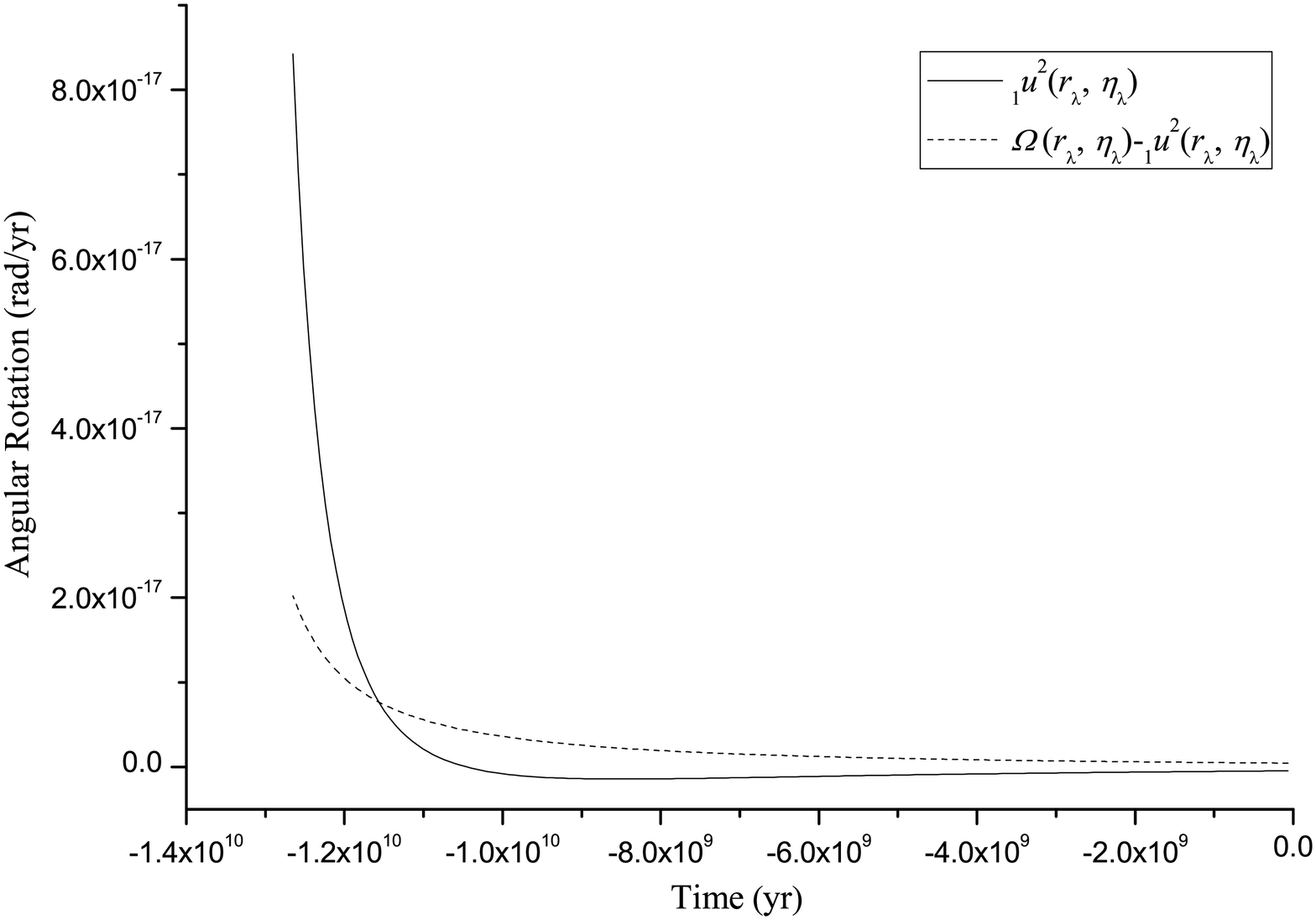}
\caption{The velocity of matter $_1u^2(r_{\lambda},\eta_{\lambda})$ and its difference from the rotation speed of the universe are plotted against time. Similar to Fig.~2, $r_{\lambda}=-\lambda\sin\phi$, $\eta_{\lambda}=\lambda$ and $\phi=\pi/2$. We note that the rotation of matter in the universe can be different and even opposite to the rotation of the universe.}
\label{fig:}
\end{figure}

\begin{figure}[hp]
\centering
\includegraphics[width=16cm, height=12cm]{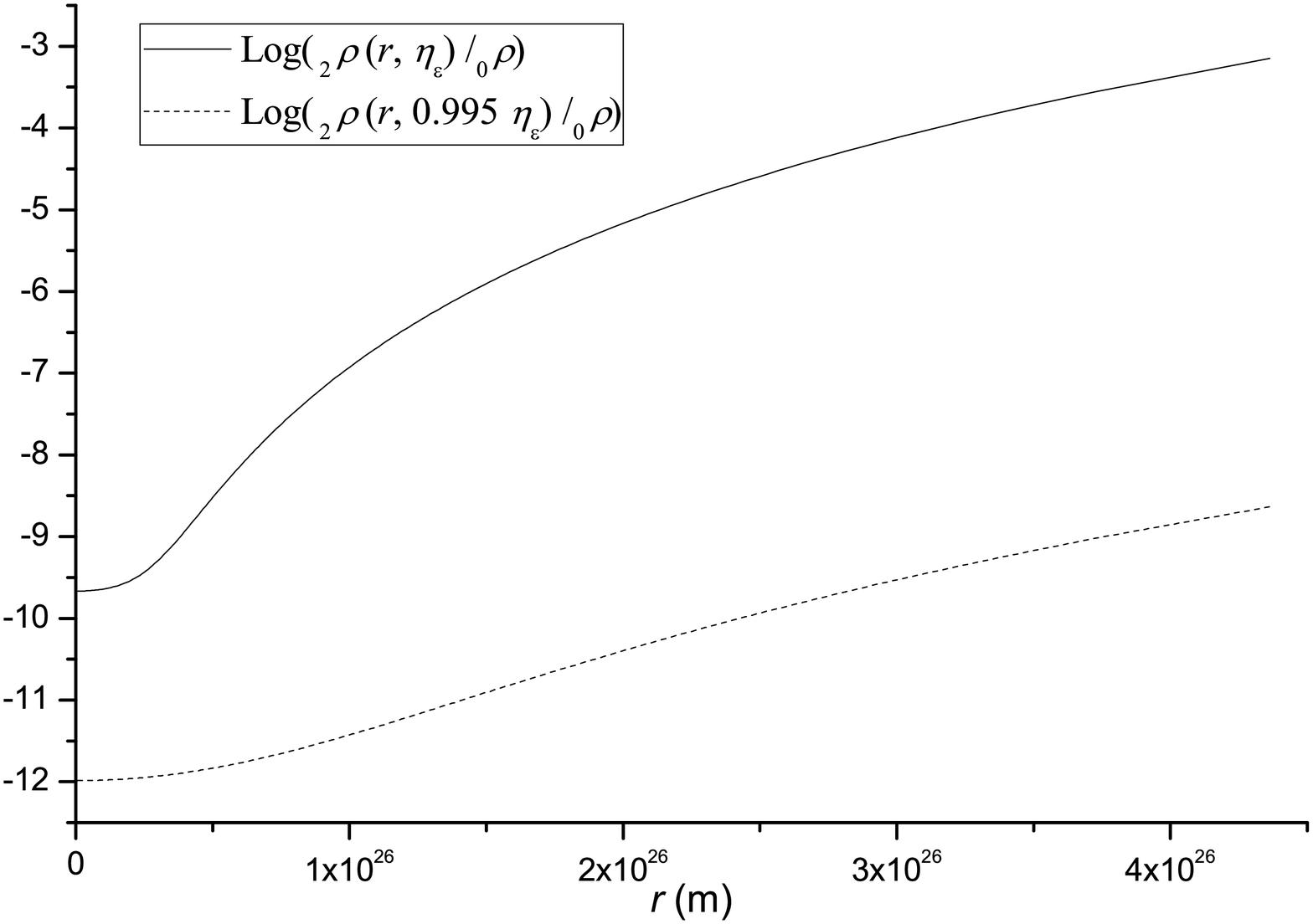}
\caption{The distributions of the 2nd-order perturbed densities of matter are shown at two different time. The perturbed density is decreasing with r, which is consistent with the positive 2nd-order radial velocity $_2u^1(r,\eta)$.}
\label{fig:}
\end{figure}

\section{Discussion and Conclusions}

In this paper, we have developed a cosmological model that has a non-zero rotation but trivial shear in the Robertson-Walker framework with a cosmological constant. We have solved the EFE's up to 2nd-order perturbations of a flat $\Lambda$CDM universe with rotation as a 1st-order perturbation. We also set up the formulation for the 2nd-order SW effect due to the rotational perturbation and find that the effect only influences the spherical harmonics with even $l$'s. By making use of recent CMBA data, the angular speed of the rotation is constrained to be less than $10^{-9}$ rad yr$^{-1}$ at the last scattering surface. 

The model of the universe here is different from the Bianchi models used in the literatures. First of all, our model is inhomogeneous with an axial symmetry in general while Bianchi models are homogeneous. Moreover, our model is shear-free and thus has the advantage that the SW effect and the constraint obtained are purely due to the global rotation.

Compared with previous works, the constraint is much weaker. For example, Barrow et al. \cite{3} put a constraint of $1.5\times 10^{-15}$ rad yr$^{-1}$ on the rotaion of flat Bianchi models. This can be understood mainly because the effects of rotation in our model here show up as 2nd-order SW effects while in previous works they are 1st-order SW effects. For further study, we notice that our model here produces a 2nd-order outward radial velocity. It may be used to explain parts of the accelerating expansion of the universe, and therefore, constraints on the global rotation can be obtained with Type Ia supernova data as proposed in \cite{SNI}. Although we only study a flat universe here, it is interesting to study the closed and open cases of our model in view of the significantly different constraints on closed and open Bianchi models\cite{3,Barrow3}.

The constraint obtained here, which is less than $10^{-9}$ rad yr$^{-1}$ at the last scattering surface, provides the first constraint of the shear-free rotation of a $\Lambda$CDM universe.

\begin{center}
   {\bf APPENDIX I}
\end{center}
Since the pressure is the same along different directions at one point, Eqs.~(19)-(21) are equivalent. By substracting Eqs.~(19)-(21) from each other, we have 
\renewcommand{\theequation}{A.\arabic{equation}}
\setcounter{equation}{0}
\begin{eqnarray}
T'(r,\eta)-rT''(r,\eta)&=&\frac{-a^2(\eta)}{8[-2\dot{a}^2(\eta)+a(\eta)\ddot{a}(\eta)]^2}
\{64r^3\dot{a}^4(\eta)\Omega '^2(r,\eta)
+16r^3a^2(\eta)\ddot{a}^2(\eta)\Omega '^2(r,\eta)\nonumber\\
&~&+r\Lambda a^6(\eta)[3\Omega '(r,\eta)+r\Omega ''(r,\eta)]^2
+ra(\eta)\dot{a}^2(\eta)\{-64r^2\ddot{a}(\eta)\Omega '^2(r,\eta)\nonumber\\
&~&+3a(\eta)[3\Omega '(r,\eta)+r\Omega ''(r,\eta)]^2\}\},\\
0&=&2r\dot{a}^2(\eta)L(r,\eta)+ra(\eta)[-4\ddot{a}(\eta)L(r,\eta)+
2\dot{a}(\eta)\dot{L}(r,\eta)]\nonumber\\
&~&+a^4(\eta)[-2r\Lambda L(r,\eta)
+r^3\Omega '^2(r,\eta)]+a^2(\eta)[r\ddot{L}(r,\eta)-L'(r,\eta)\nonumber\\
&~&-rL''(r,\eta)-rT''(r,\eta)].
\end{eqnarray}

These two equations are independent of $k(r,\eta)$. That is, $k(r,\eta)$ is only an arbitrary function unrelated to $\Omega(r,\eta)$ in general. Physically, $k(r,\eta)$ comes from the 2nd-order perturbation of mass density (analogous to the diagonal perturbations of Schwarzschild metric). As we are interested in the effects of the rotation only, we set $k(r,\eta)=0$ for simplicity. 

In this paper, we focus on a non-viscous fluid in the universe. Using the fact that $_0\rho(\eta)=\Lambda(1-\Omega_\Lambda)/[8\pi \Omega_\Lambda a^3(\eta)]$ for a flat $\Lambda$CDM universe and Eq.~(8), we can simplify these two equations further
\begin{eqnarray}
T'(r,\eta)-rT''(r,\eta)=-\frac{2r^3A'^2(r)}{a^4(\eta)}-
\frac{\Omega_{\Lambda}r[3A'(r)+rA''(r)]^2}{2\Lambda(1-\Omega_{\Lambda})a^3(\eta)},\\
\frac{r}{a(\eta)}\frac{d}{d\eta}[a^2(\eta)\dot{L}(r,\eta)]-a(\eta)[L'(r,\eta)
+rL''(r,\eta)]=ra(\eta)T''(r,\eta)-r^3a^{-3}(\eta)A'^2(r).
\end{eqnarray}

Here, we remark that $B(\eta)$, which disappears from these two equations, is arbitrary and does not affect the Sachs-Wolfe effect. Such an arbitrariness is  unrelated to the effects of the rotational universe on the CMBA.

Without loss of generality, we expand $A(r)$ into Taylor series: $A(r)=\Omega_{rot}\sum_{n=2}^{\infty}c_nr^n$ (we will explain below why it does not start from $n=0$). Substituting the series into Eq.~(A.3), 
\begin{eqnarray}
&~&T'(r,\eta)-rT''(r,\eta)\nonumber\\
&=&-\frac{2\Omega_{rot}^2}{a^4(\eta)}\sum_{n=5}^{\infty}\sum_{l=2}^{\infty}
l(n-l-1)c_lc_{n-l-1}r^n\nonumber\\
&~&-\frac{\Omega_{\Lambda}\Omega_{rot}^2}{2\Lambda(1-\Omega_{\Lambda})a^3(\eta)}\sum_{n=3}^{\infty}\sum_{l=2}^{\infty}
l(l+2)(n-l+1)(n-l+3)c_lc_{n-l+1}r^n,
\end{eqnarray}
where $c_n=0$ for $n<2$.

By separation of variables, we find that
\begin{eqnarray}
T(r,\eta)&=&\Omega_{rot}^2\left[D_1(\eta)r^2+D_2(\eta)
-\frac{2}{a^4(\eta)}\xi_1(r)-\frac{\Omega_{\Lambda}}{2\Lambda(1-\Omega_{\Lambda})a^3(\eta)}\xi_2(r)\right],
\end{eqnarray}
where
\begin{eqnarray}
\xi_1(r)&=&\sum_{n=5}^{\infty}\sum_{l=2}^{\infty}
\frac{l(n-l-1)}{(n+1)(n-1)}c_lc_{n-l-1}r^{n+1},\\
\xi_2(r)&=&\sum_{n=3}^{\infty}\sum_{l=2}^{\infty}
\frac{l(l+2)(n-l+1)(n-l+3)}{(n+1)(n-1)}c_lc_{n-l+1}r^{n+1}.
\end{eqnarray}
As $D_1(\eta)$ and $D_2(\eta)$ are arbitrary functions of the homogeneous solutions for Eq.~(A.5) and thus are independent of $\Omega(r,\eta)$, we are free to set them zero. The series of $A(r)$ starts from $n=2$ because the $n=0$ term can be absorbed into $B(\eta)$ while the $n=1$ term, which produces $\ln(r)$ as the particular solution of $T(r,\eta)$, is rejected because of the singularity at $r=0$.

Similarly, we expand $L(r,\eta)=\Omega_{rot}^2\sum_{n=0}^{\infty}E_n(\eta)r^n$ and substitute it into Eq.~(A.4):
\begin{eqnarray}
&~&\sum_{n=0}^{\infty}\left\{\frac{1}{a(\eta)}\frac{d}{d\eta}[a^2(\eta)\dot{E}_{n-1}(\eta)]
-(n+1)^2a(\eta)E_{n+1}(\eta)\right\}r^n\nonumber\\
&=&-\sum_{n=5}^{\infty}\sum_{l=2}^{\infty}\frac{l(3n-1)(n-l-1)}{(n-1)a^3(\eta)}
c_lc_{n-l-1}r^{n}\nonumber\\
&~&-\sum_{n=3}^{\infty}\sum_{l=2}^{\infty}
\frac{nl(l+2)(n-l+1)(n-l+3)\Omega_{\Lambda}}{2\Lambda(1-\Omega_{\Lambda})(n-1) a^2(\eta)}c_lc_{n-l+1}r^{n},
\end{eqnarray}
where $E_n(\eta)=0$ for $n<0$. In general, by comparing the $r^n$ terms on both sides, we can obtain the recurrence relations for solving $E_n(\eta)$. As an example, we work out the simplest case where $\Omega(r,\eta)=\alpha\Omega_{rot} r^2a^{-3}(\eta)$ ($\alpha$ is a constant): 
\begin{eqnarray}
T(r,\eta)&=&-\frac{\alpha^2\Omega_{rot}^2}{3a^4(\eta)}r^6
-\frac{4\Omega_{\Lambda}\alpha^2\Omega_{rot}^2}{\Lambda(1-\Omega_{\Lambda})a^3(\eta)}r^4,\\
0&=&E_{2n+1}(\eta),\\
0&=&\frac{1}{a(\eta)}\frac{d}{d\eta}[a^2(\eta)\dot{E}_0(\eta)]-4a(\eta)E_2(\eta),\\
-\frac{48\Omega_{\Lambda}\alpha^2}{\Lambda(1-\Omega_{\Lambda})a^2(\eta)}&=&\frac{1}{a(\eta)}
\frac{d}{d\eta}[a^2(\eta)\dot{E}_2(\eta)]-16a(\eta)E_4(\eta),\\
-\frac{14\alpha^2}{a^3(\eta)}&=&\frac{1}{a(\eta)}
\frac{d}{d\eta}[a^2(\eta)\dot{E}_4(\eta)]-36a(\eta)E_6(\eta),\\
&\vdots&\nonumber\\
0&=&\frac{1}{a(\eta)}\frac{d}{d\eta}[a^2(\eta)\dot{E}_{2n-2}(\eta)]-4n^2a(\eta)E_{2n}(\eta).
\end{eqnarray}

We notice that there is a freedom to choose one of $E_{2n}(\eta)$ arbitrarily, independent of $\Omega(r,\eta)$. To prevent infinite series , we set $E_{6}(\eta)=0$ so that $E_{2n}(\eta)=0$ for $n>3$ and all non-zero $E_{2n}(\eta)$'s depend on $\alpha^2$ (due to the rotation). We have
\begin{eqnarray}
E_4(\eta)&=&\int_{\eta}^0\frac{1}{a^2(\eta ')}\int_{\eta '}^0
\frac{-14\alpha^2}{a^2(\eta '')}d\eta ''d\eta ',\\
E_2(\eta)&=&\int_{\eta}^0\frac{1}{a^2(\eta ')}\int_{\eta '}^0\left[\frac
{-48\Omega_{\Lambda}\alpha^2}{\Lambda(1-\Omega_{\Lambda})a(\eta '')}+16a^2(\eta '')E_4(\eta '')\right]d\eta ''d\eta ',\\
E_0(\eta)&=&\int_{\eta}^0\frac{1}{a^2(\eta ')}\int_{\eta '}^0
4a^2(\eta '')E_2(\eta '')d\eta ''d\eta ',
\end{eqnarray}
which can be solved numerically.

\begin{center}
   {\bf APPENDIX II}
\end{center}
Using Eq.~(22) and expanding $u^{\mu}$, $k^{\nu}$ up to 2nd-order in $\Omega_{rot}$ as in Eqs.~(4)-(7), we have the temperature anisotropy
\renewcommand{\theequation}{B.\arabic{equation}}
\setcounter{equation}{0}
\begin{eqnarray}
\frac{\delta T_{\epsilon}}{T_{\epsilon}}
&=&(_0k^{\mu}\,_1u_{\mu}+{_1k^{\mu}}\,_0u_{\mu})|_{\eta_{\epsilon}}^{\eta_0}-
(_0k^{\mu}\,_1u_{\mu}+{_1k^{\mu}}\,_0u_{\mu})|_{\eta_{\epsilon}}
(_0k^{\mu}\,_1u_{\mu}+{_1k^{\mu}}\,_0u_{\mu})|_{\eta_0}\nonumber\\
&~&+[(_0k^{\mu}\,_1u_{\mu}+{_1k^{\mu}}\,_0u_{\mu})|_{\eta_{\epsilon}}]^2+
(_0k^{\mu}\,_2u_{\mu}+{_1k^{\mu}}\,_1u_{\mu}+{_2k^{\mu}}\,_0u_{\mu})
|_{\eta_{\epsilon}}^{\eta_0}.
\end{eqnarray}
In order to calculate Eq.~(B.1), we need to solve the geodesic equations for the light rays of the CMB, which are
\begin{eqnarray}
\frac{d}{d\lambda}\left(g_{\mu\nu}\frac{dx^{\nu}}{d\lambda}\right)=
\frac{1}{2}g_{\nu\gamma,\mu}\frac{dx^{\nu}}{d\lambda}\frac{dx^{\gamma}}{d\lambda}.
\end{eqnarray}
As before, we expand the geodesic equations into different orders of $\Omega(r,\eta)$. To be consistent, $g_{\mu\nu}(x^{\gamma})$ has to be expanded into
\begin{eqnarray}
g_{\mu\nu}(x^{\gamma})&=&g_{\mu\nu}(_0x^{\gamma}+{_1x^{\gamma}}+{_2x^{\gamma}})\nonumber\\
&=&g_{\mu\nu}(_0x^{\gamma})+{_1x^{\gamma}}g_{\mu\nu,\gamma}(_0x^{\gamma})
+{_2x^{\gamma}}g_{\mu\nu,\gamma}(_0x^{\gamma})
+\frac{_1x^{\gamma}_1x^{\zeta}}{2}g_{\mu\nu,\gamma\zeta}(_0x^{\gamma})
+O(\Omega^3).
\end{eqnarray}

For the zeroth-order, it is trivial that 
\begin{eqnarray}
_0k^{\mu}=(1,-\sin\phi,0,-\cos\phi),
\end{eqnarray}
where $\phi$ is the zenith angle measured from the $z$-axis.

For the first-order, 
\begin{eqnarray}
\frac{d}{d\lambda}\left[_0g_{22}\frac{d_1x^2}{d\lambda}+{_1g_{20}}\frac{d_0x^0}{d\lambda}\right]=0,\\
\frac{d}{d\lambda}\left[_0g_{ii}\frac{d_1x^i}{d\lambda}\right]=0,
\end{eqnarray}
for $i=0,1,3$. Therefore, $_1k^{\mu}(\lambda)=(A_0,A_1,a(\eta_{\lambda})\Omega(r_{\lambda},\eta_{\lambda})
-A_2/r_{\lambda}^2,A_3)$ where $A_n$'s are constants to be determined. For simplicity, we assume that we are located on the rotating axis. Therefore, $\eta_{\lambda}=\lambda$ and $r_{\lambda}=-\lambda\sin\phi$ due to the cylindrical symmetry. Although the general case that we may be off the rotating axis is more realistic, the constraint here can be regarded as a good approximation provided that our distance to the rotating axis is small compared to that of the last scattering surface. The general case can be found by assigning a suitable dependence of $\theta$ on $\eta_{\lambda}$ and $r_{\lambda}$. The term $a(\eta_{\lambda})\Omega(r_{\lambda},\eta_{\lambda})$ refers to the angular velocity of the comoving metric.

For the second-order,
\begin{eqnarray}
&~&\frac{d}{d\lambda}\left(_0g_{00}\frac{d_2x^0}{d\lambda}+
{_1g_{02}}\frac{d_1x^2}{d\lambda}+{_2g_{00}}\frac{d_0x^0}{d\lambda}\right)\nonumber\\
&=&_1g_{20,0}\frac{d_1x^2}{d\lambda}\frac{d_0x^0}{d\lambda}+
\frac{_2g_{00,0}}{2}\left(\frac{d_0x^0}{d\lambda}\right)^2+
\frac{_2g_{11,0}}{2}\left(\frac{d_0x^1}{d\lambda}\right)^2+
\frac{_2g_{33,0}}{2}\left(\frac{d_0x^3}{d\lambda}\right)^2,\\
&~&\frac{d}{d\lambda}\left(_0g_{11}\frac{d_2x^1}{d\lambda}+{_2g_{11}}\frac{d_0x^1}{d\lambda}\right)\nonumber\\
&=&\frac{_0g_{22,1}}{2}\left(\frac{d_1x^2}{d\lambda}\right)^2+
{_1g_{20,1}}\frac{d_1x^2}{d\lambda}\frac{d_0x^0}{d\lambda}+
\frac{_2g_{00,1}}{2}\left(\frac{d_0x^0}{d\lambda}\right)^2+
\frac{_2g_{11,1}}{2}\left(\frac{d_0x^1}{d\lambda}\right)^2+\nonumber\\
&~&\frac{_2g_{33,1}}{2}\left(\frac{d_0x^3}{d\lambda}\right)^2,\\
0&=&\frac{d}{d\lambda}\left(_0g_{22}\frac{d_2x^2}{d\lambda}+{_1x^1}\,_0g_{22,1}\frac{d_1x^2}{d\lambda}+
{_1g_{20}}\frac{d_1x^0}{d\lambda}+{_1x^0}\,_1g_{20,0}\frac{d_0x^0}{d\lambda}+
{_1x^1}\,_1g_{20,1}\frac{d_0x^0}{d\lambda}\right),\\
0&=&\frac{d}{d\lambda}\left(_0g_{33}\frac{d_2x^3}{d\lambda}+{_2g_{33}}\frac{d_0x^3}{d\lambda}\right).
\end{eqnarray}
By setting $k(r,\eta)=0$ and the requirement of a null geodesic ($k^{\mu}k_{\mu}=0$), we obtain
\begin{eqnarray}
_1k^{\mu}(\lambda)&=&(0,0,a(\eta_{\lambda})\Omega(r_{\lambda},\eta_{\lambda}),0),\\
_2k^{\mu}(\lambda)&=&(_2k^0(\lambda),{_2k^1(\lambda)},0,0),
\end{eqnarray}
where
\begin{eqnarray}
_2k^0(\lambda)&=&\int\left[\frac{\dot{T}(r_{\lambda '},\eta_{\lambda '})}{2}
-T'(r_{\lambda '},\eta_{\lambda '})\sin\phi+
\frac{\dot{L}(r_{\lambda '},\eta_{\lambda '})}
{2}\sin^2\phi\right]\,d\lambda ',\\
_2k^1(\lambda)&=&\int\left[\frac{T'(r_{\lambda '},\eta_{\lambda '})}{2}+
\frac{L'(r_{\lambda '},\eta_{\lambda '})}{2}\sin^2\phi
-\dot{L}(r_{\lambda '},\eta_{\lambda '})\sin\phi\right]\,d\lambda '.
\end{eqnarray}
To have a null geodesic, 
\begin{eqnarray}
0&=&2_2k^0(\lambda)+T(r_{\lambda},\eta_{\lambda})
+2_2k^1(\lambda)\sin\phi-L(r_{\lambda},\eta_{\lambda})\sin^2\phi,
\end{eqnarray}
for $\lambda=0$.
From the results of $k^{\mu}$ and $u_{\mu}$, we can easily verify the argument that the 1st-order perturbations of SW effect due to the rotation is zero and calculate the 2nd-order SW effect as
\begin{eqnarray}
&~&\frac{\delta T_{\epsilon}}{T_{\epsilon}}
=\left[-\frac{_2u_1(r_{\lambda},\eta_{\lambda})}{a(\eta_{\lambda})}\sin\phi
+\frac{_2u_0(r_{\lambda},\eta_{\lambda})}{a(\eta_{\lambda})}+_2k^0(\lambda)
+\frac{_1u_2(r_{\lambda},\eta_{\lambda})}{a(\eta_{\lambda})}{_1k^2(\lambda)}
\right]|_{\eta_{\epsilon}}^{\eta_0}\nonumber\\
&=&\frac{\Omega_{\Lambda}\sin\phi}{2\Lambda(1-\Omega_{\Lambda})}
[2\dot{a}(\eta_{\epsilon})T'(r_{\epsilon},\eta_{\epsilon})-
a(\eta_{\epsilon})\dot{L}'(r_{\epsilon},\eta_{\epsilon})]\nonumber\\
&~&-\frac{1}{2}\left\{\frac{\Omega_{\Lambda}^2a^4(\eta_{\epsilon})}{4\Lambda^2(1-\Omega_{\Lambda})^2}
[3\Omega '(r_{\epsilon},\eta_{\epsilon})+r_{\epsilon}\Omega ''(r_{\epsilon},\eta_{\epsilon})]^2+T(r_{\epsilon},\eta_{\epsilon})\right\}\nonumber\\
&~&+\int_{\eta_{\epsilon}}^0\left[-\frac{\dot{T}(-\lambda\sin\phi,\lambda)}{2}+
T'(-\lambda\sin\phi,\lambda)\sin\phi-\frac{\dot{L}(-\lambda\sin\phi,\lambda)}{2}\sin^2\phi\right]d\lambda,
\end{eqnarray}
where $r_{\epsilon}=-\eta_{\epsilon}\sin\phi$ and $\eta_{\epsilon}$ denotes the occuring conformal time of the last scattering.

\begin{acknowledgments}

We made use of HEALPIX \cite{healpix} to produce Fig. 1. This work is supported by grants from the Research Grant Council of the Hong Kong Special Administrative Region, China (Project Nos. 400707 and 400803).

\end{acknowledgments}

\end{document}